\documentclass[prb,twocolumn,showpacs,superscriptaddress,floatfix,amsmath]{revtex4}
\usepackage{graphicx}
\makeatother

\begin{document}
\title{Embedding approach for dynamical mean field theory of strongly correlated
heterostructures}
\author{H. Ishida}
\affiliation{College of Humanities and Sciences, Nihon University, Tokyo
156-8550, Japan}
\author{A. Liebsch}
\affiliation{Institut f\"ur Festk\"orperforschung, Forschungszentrum J\"ulich,
52425 J\"ulich, Germany}
\date{\today}
\begin{abstract}
We present an embedding approach based on localized basis functions which permits 
an efficient application of the dynamical mean field theory (DMFT) to inhomogeneous 
correlated materials, such as semi-infinite surfaces and heterostructures.
In this scheme, the semi-infinite substrate leads connected to both sides of 
the central region of interest are represented via complex, energy-dependent 
embedding potentials that 
incorporate one-electron as well as many-body effects within the substrates.
As a result, the number of layers which must be treated explicitly in the
layer-coupled DMFT equation is greatly reduced. To illustrate the usefulness of 
this approach, we present numerical results for strongly correlated surfaces, 
interfaces, and heterostructures of the single-band Hubbard model.
\end{abstract}

\maketitle

\section{Introduction}

In recent years, there is growing interest in the electronic properties
of surfaces and interfaces of strongly correlated materials.\cite{Dagotto:07}
For instance, the discrepancy between photoemission spectra and theoretically
derived bulk spectra of a number of transition-metal oxides has been
attributed to changes in the electronic structure at the surface of these
materials.\cite{Matzdorf:00,Maiti:01,Sekiyama:04}
Regarding the interface, heterostructures made out of thin atomic layers
of perovskite-type oxides have been the target of intense study as 
promising candidates for electron-correlation-based
devices.\cite{Ohtomo:02,Takizawa:06,Kourkoutis:06,Maekawa:07,Wadati:08}
A well-known example is the interface between LaTiO$_3$ and
SrTiO$_3$, which exhibits metallic behavior in spite of the fact that the
two constituent bulk materials are insulators.\cite{Ohtomo:02}

On the theoretical side, inhomogeneous layered systems have been studied
by several authors within the dynamical mean field theory\cite{Georges:96} (DMFT), 
which contributed significantly in the last decade to the understanding of a 
variety of strongly correlated bulk materials.\cite{recent.dmft}
Pott\-hoff and Nolting investigated the 
metal-insulator transition (MIT) at the surface of the single-band Hubbard
model.\cite{Potthoff:99,Potthoff:01}
{Liebsch} studied the $t_{2g}$ valence bands of perovskite-type oxides such as
SrVO$_3$ by using a three-band tight-binding Hamiltonian and showed that
electrons at the surface are more strongly correlated than in the bulk due to
the reduction of the effective surface band 
width.\cite{Liebsch:03}
Helmes {\it et al.} considered a metal-insulator interface within the single-band
Hubbard model and studied the scaling behavior of the metallic penetration depth
into the Mott insulator near the critical Coulomb energy.\cite{Helmes:08}
Okamoto and Millis investigated the electronic structure of heterostructures
in which a finite number of Mott-insulator layers were sandwiched between
band insulators.\cite{Okamoto:04,Okamoto:05}
Analogous calculations for heterostructures consisting of correlated model
systems were also carried out by Kancharla and Dagotto\cite{Kancharla:06} and
R\"uegg {\it et al.}\cite{Ruegg:07}
Modulation doping effects at heterojunctions were investigated by 
Oka and Nagaosa\cite{Oka:05}, 
Lee and MacDonald\cite{Lee:06} and
Gonz\'alez {\it et al.}\cite{Gonzalez:08}
Electron transport through a nano-size correlated-electron system connected
to metal electrodes was studied by combining DMFT and a non-equilibrium
Green-function technique.\cite{Ferretti:05,Okamoto:08}

To solve the DMFT equation for inhomogeneous layered systems one needs to
construct the lattice Green function of surfaces or interfaces consisting of
an infinite number of atomic layers. While this is feasible within a
linearized version of DMFT,\cite{Potthoff:99}
for a complete numerical solution of the DMFT equation most previous
calculations employed a slab model consisting of a finite number of layers
to simulate the system. Although finite-size effects can be reduced by
systematically increasing the number of layers, the one-electron density of
states (DOS) projected on each layer converges rather slowly with increasing
number of layers since the energy levels in the normal direction are
discrete. Hence, it is desirable to develop a method for solving the DMFT
equation for truly semi-infinite surfaces and interfaces between two
semi-infinite materials.
Chen and Freericks solved the DMFT equation for a thin doped Mott insulator
sandwiched between two semi-infinite metals by applying the quantum zipper
algorithm to the Falikov-Kimball Hamiltonian.\cite{Chen:07}
In the present work we pursue a different approach by extending the concept 
of tight-binding embedding, originally developed for the evaluation of the
electronic properties of defects in solids, to the DMFT for 
inhomogeneous layered systems.

We employ a localized basis set to describe the Hamiltonian of the system.
The heterostructure is divided into a central interface region containing a 
finite number of atomic layers, $\Omega$, and two adjacent semi-infinite 
bulk regions coupled to $\Omega$. The interface region is assumed to include 
the first few surface layers of the actual substrates. Both the central region 
and the substrates may exhibit strong correlation effects.  
Within the one-electron approximation, 
the effects of an adjacent semi-infinite system on $\Omega$ can be expressed 
as a complex energy-dependent potential acting on the Hamiltonian matrix of 
$\Omega$, which is called ``tight-binding embedding
potential''.\cite{Inglesfield:01,Baraff:86,Pisani:79}
The same quantity is called ``contact self-energy'' in transport theory based
on the non-equilibrium Green function formalism.\cite{Datta:95,Thygesen:08}
Here, we extend this embedding approach in order to include Coulomb correlations
in the substrate within the single-site DMFT. Thus, the energy-dependent
embedding potential accounts for one-electron and many-body effects within 
the substrates. The advantage of this extension is that the layer-coupled DMFT
equation for a non-periodic surface or interface system made up of an infinite
number of atomic layers is greatly simplified since only a small number of
layers belonging to $\Omega$ needs to be treated explicitly. The embedding 
potential is derived from a separate DMFT calculation for the adjacent bulk 
systems. 

The plan of this paper is as follows: In Sec.\ \ref{sec_2} we present our
formalism that combines the tight-binding embedding theory with DMFT.
As examples, in Sec.\ \ref{sec_3} we present numerical results for surfaces
and interfaces of the single-band Hubbard model at half filling. A summary
is given in Sec.~\ref{sec_4}.

\section{Theory}
\label{sec_2}

\subsection{Hamiltonian}

We take the $z$ axis as the surface normal pointing from left to right. The
$j^{th}$ atomic layer is located at $z=z_j$ ($z_j>z_{j-1}$). The position of
the $m^{th}$ atom in layer $j$ is denoted by ${\bf r}_a=({\bf x}_{mj}, z_j)$,
where the index $a$ represents a pair of indices
$(m,j)$. The localized basis function centered at ${\bf r}_a$ with orbital
index $\beta$ and spin index $\sigma$ is denoted by $\psi_{a \beta \sigma}$.
The basis set
$\lbrace\psi_{a \beta \sigma}\rbrace$ is assumed to be orthonormal.
Hereafter, we use indices with tilde such as $\tilde{p}$ and $\tilde{q}$ to
refer to basis functions in the basis set
$\lbrace\psi_{a \beta \sigma}\rbrace$. With this abbreviated
notation, the one-electron part of the Hamiltonian is written as
\begin{equation}
\label{eq1}
\hat{h}=\sum_{\tilde{p}, \tilde{q}} t_{\tilde{p}\tilde{q}}\
c^\dagger_{\tilde{p}} c_{\tilde{q}},
\end{equation}
where $c^\dagger_{\tilde{p}}$ and $c_{\tilde{q}}$ are the creation and
annihilation operators, respectively, and
summation is taken over pairs $(\tilde{p},\tilde{q})$ having the same
spin and located on the same or nearby atomic sites.
The one-electron Hamiltonian may be derived, for example,
from a first-principles electronic-structure calculation within
density-functional theory through the use of maximally localized Wannier
functions.\cite{Marzari:97}

In the present work we consider onsite Coulomb interactions,
\begin{equation}
\label{eq2}
\hat{v}=\frac{1}{2} \sum_{\tilde{p},\tilde{q},  \tilde{r}, \tilde{s}}
V_{\tilde{p}\tilde{q},\tilde{r}\tilde{s}} c^\dagger_{\tilde{p}}
c^\dagger_{\tilde{q}} c_{\tilde{s}} c_{\tilde{r}},
\end{equation}
where $\tilde{p}$, $\tilde{q}$, $\tilde{r}$, and $\tilde{s}$ are located
on the same site, and in addition, $\tilde{p}$ and $\tilde{r}$
($\tilde{q}$ and $\tilde{s}$) have the same spin.
The full Hamitonian of the system is given by
$\hat{H}=\hat{h}+\hat{v}$.

We now divide the system into three parts. The central region with atomic
layer index $j$ running from 1 to $N$ is called $\Omega$. The semi-infinite
region with layer number $j < 1$ is called ``left substrate'' $L$,
and the semi-infinite region with $j >N$ is called ``right substrate'' $R$.
In the case of a semi-infinite surface, it is understood that the system
consists only of $\Omega$ and $R$. In the following, we present the theory 
for the interface geometry. The analogous equations theory for a semi-infinite 
surface are derived straightforwardly by omitting all terms with
index $L$. The one-electron Hamiltonian $\hat{h}$ in Eq.\ (\ref{eq1}) is
decomposed into seven parts:
\begin{equation}
\label{eq3}
\hat{h}=\hat{h}_{LL}+\hat{h}_{\Omega\Omega}+\hat{h}_{RR}
+\sum_{J=L,R} \left(\hat{h}_{\Omega J}
+\hat{h}_{J\Omega}\right),
\end{equation}
with
\begin{equation}
\label{eq4}
\hat{h}_{JK}=\sum_{\tilde{j},\tilde{k}} \langle\tilde{j}|\hat{h}|
\tilde{k}\rangle c^\dagger_{\tilde{j}} c_{\tilde{k}},
\end{equation}
where $J$ and $K$ denote one of the three regions,  $L$,
$\Omega$, and $R$, and the basis function $\tilde{j}$ ($\tilde{k}$) belongs
to region $J$ ($K$). It is to be noted that the matrix elements of
the inter-regional terms, $t_{\tilde{w}\tilde{j}}$,
where  $\tilde{w}\in\Omega$ and $\tilde{j}\in J$ ($J=L, R$), are
non-vanishing only when $\tilde{w}$ and $\tilde{j}$ are close
to the boundary between $\Omega$ and $J$, since transfer integrals are
short-ranged. For the same reason, $\hat{h}_{LR}$ and $\hat{h}_{RL}$
are assumed to vanish.

\subsection{Non-interacting Green function}

As a brief review of the tight-binding embedding
theory,\cite{Inglesfield:01,Baraff:86,Pisani:79} we outline first the calculation 
of the Green function (resolvent) of the one-electron Hamiltonian $\hat{h}$,
\begin{equation}
\label{eq5}
g_{\tilde{p}\tilde{q}}(\epsilon) =
\langle\tilde{p}|(\epsilon-\hat{h})^{-1}|\tilde{q}\rangle.
\end{equation}
When both indices of this Green function belong to $\Omega$, the tight-binding 
embedding theory reveals that the interaction with the left and right substrates 
can be expressed in terms of embedding potentials acting on $\Omega$:
\begin{eqnarray}
      s_{\tilde{w}\tilde{w}'}^L(\epsilon) &=&  t_{\tilde{w}\tilde{l}}\
g_{\tilde{l}\tilde{l}'}^L(\epsilon)\ t_{\tilde{l}'\tilde{w}'}, \label{eq6a}\\
      s_{\tilde{w}\tilde{w}'}^R(\epsilon) &=&  t_{\tilde{w}\tilde{r}}\
g_{\tilde{r}\tilde{r}'}^R(\epsilon)\ t_{\tilde{r}'\tilde{w}'}, \label{eq6b}
\end{eqnarray}
where $\tilde{w}, \tilde{w}'\in \Omega$ and the summation is implied for
repeated indices. $\hat{g}^L$ and $\hat{g}^R$ are the Green functions
of the left and right substrates, respectively, and are defined as
\begin{eqnarray}
\hat{g}^L(\epsilon) &=& (\epsilon-\hat{h}_{LL})^{-1}, \label{eq7a}\\
\hat{g}^R(\epsilon) &=& (\epsilon-\hat{h}_{RR})^{-1}.\label{eq7b}
\end{eqnarray}
It should be noted that  $\hat{     s}^L$ and $\hat{     s}^R$ in
Eqs.\ (\ref{eq6a}) and (\ref{eq6b}) are non-vanishing only when both
$\tilde{w}$ and $\tilde{w}'$ are located close to the boundaries of $\Omega$.

Using these embedding potentials, the Green function defined in
Eq.\ (\ref{eq5}), when both indices belong to $\Omega$, can be calculated as
\begin{equation}
\label{eq7c}
g_{\tilde{w}\tilde{w}'}(\epsilon) =\langle\tilde{w}\mid
[\epsilon-\hat{h}_{\rm em}(\epsilon)]^{-1}\mid\tilde{w}'\rangle,
\end{equation}
where the effective Hamiltonian in the embedded region $\Omega$ is given by
\begin{equation}
\label{eq7d}
\hat{h}_{\rm em}=\hat{h}_{\Omega\Omega}
 +\hat{     s}^L(\epsilon)+\hat{     s}^R(\epsilon).
\end{equation}
Thus, the calculation of a system which is non-periodic in the $z$
direction is reduced to the inversion of a matrix defined in $\Omega$
with a finite thickness. 

We point out that, inspite of the finite size 
of the central interface region, the use of the complex embedding potentials
ensures that the spectral distribution is continuous. In particular,
for a uniform system with layer-independent energy levels and hopping
matrix elements, the local DOS of each layer coincides
with the bulk DOS. Thus, there are no discretization effects 
stemming from the finite number of layers in the central region.    

\subsection{Dynamical mean field theory}

We now incorporate the Coulomb interactions and calculate the finite-temperature
Green function of the full Hamiltonian, $\hat{H}=\hat{h}+\hat{v}$.
The effects of the Coulomb interactions can be described by a
frequency-dependent self-energy
$\hat{\Sigma}(i\omega_n)$, where $\omega_n=(2n+1)\pi T$ are
Matsubara frequencies at temperature $T$. In the present work, we restrict
ourselves to the single-site approximation and assume that the matrix
elements
$\Sigma_{\tilde{p}\tilde{q}}(i\omega_n)$ are non-vanishing
only when $\tilde{p}$ and $\tilde{q}$ are on the same atomic site. 
Hence, 
\begin{equation}
\label{eq8}
 \hat{\Sigma}(i\omega_n)=\sum_{J,K}\hat{\Sigma}_{JK}(i\omega_n)\delta_{JK},
\end{equation}
where $\hat{\Sigma}_{JK}$ is defined in the same way as
Eq.\ (\ref{eq4}) with $\hat{h}$ replaced by $\hat{\Sigma}$.
The lattice Green function of the whole system is defined by
\begin{equation}
\label{eq9}
G_{\tilde{p}\tilde{q}}(i\omega_n) =
\langle\tilde{p}|[i\omega_n+\mu-\hat{h}-\hat{\Sigma}(i\omega_n)]
^{-1}|\tilde{q}\rangle,
\end{equation}
where $\mu$ denotes the chemical potential of the system.

As in the case of non-interacting systems, we define the embedding
potentials of the left and right correlated substrates as
\begin{eqnarray}
      S_{\tilde{w}\tilde{w}'}^L(i\omega_n)
&=&  t_{\tilde{w}\tilde{l}}\ G_{\tilde{l}\tilde{l}'}^L(i\omega_n)\
t_{\tilde{l}'\tilde{w}'}, \label{eq10a}\\
      S_{\tilde{w}\tilde{w}'}^R(i\omega_n)
&=&  t_{\tilde{w}\tilde{r}}\ G_{\tilde{r}\tilde{r}'}^R(i\omega_n)\
t_{\tilde{r}'\tilde{w}'}, \label{eq10b}
\end{eqnarray}
where $\hat{G}^L(i\omega_n)$ and $\hat{G}^R(i\omega_n)$ are
defined by
\begin{eqnarray}
\hat{G}^L(i\omega_n) &=&\left[i\omega_n+\mu
-\hat{h}_{LL}-\hat{\Sigma}_{LL}(i\omega_n)\right]^{-1},
\label{eq11a}\\
\hat{G}^R(i\omega_n) &=&\left[i\omega_n+\mu
-\hat{h}_{RR}-\hat{\Sigma}_{RR}(i\omega_n)\right]^{-1}.
\label{eq11b}
\end{eqnarray}
With these definitions, the lattice Green function defined by Eq.\ (\ref{eq9}),
when both indices belong to $\Omega$, is calculated as
\begin{equation}
\label{eq12}
G_{\tilde{w}\tilde{w}'}(i\omega_n) =
\langle\tilde{w}|[i\omega_n+\mu-\hat{H}_{\rm em}(i\omega_n)]
^{-1}|\tilde{w}'\rangle,
\end{equation}
where the interacting embedded Hamiltonian is given by
\begin{equation}
\label{eq13}
\hat{H}_{\rm em}=\hat{h}_{\Omega\Omega}
+\hat{\Sigma}_{\Omega\Omega}(i\omega_n)
 +\hat{     S}^L(i\omega_n)+\hat{     S}^R(i\omega_n).
\end{equation}

Suppose now that both substrates are semi-infinite crystals having
three-dimensional translational symmetry and that the boundary
between $\Omega$ and $L$ ($R$) is positioned a few atomic layers
toward the interior of the crystal such that the electronic structure
in $L$ ($R$) converges to that of the bulk. We may then assume that
the matrix elements of $\hat{\Sigma}_{JJ}$ on all atomic sites
in $J$ ($J=L,R$) become identical with those of the Coulomb
self-energy on the corresponding atomic site in the bulk crystal.
Therefore, we are left with determining the self-energy in the embedded
region, $\hat{\Sigma}_{\Omega\Omega}$. This can be achieved via the following
three steps:
(i) perform a standard DMFT calculation for the bulk crystals
corresponding to the left and right substrates to obtain the Coulomb
self-energies in the bulk,
(ii) construct the embedding potentials of both substrates,
 $\hat{     S}^L$ and $\hat{     S}^R$, and
(iii) perform a layer-coupled DMFT calculation in the embedded region $\Omega$
to self-consistently determine $\hat{\Sigma}_{\Omega\Omega}$.

The embedded DMFT calculation in the third step is conducted in a standard
manner. Starting with an input lattice self-energy
$\hat{\Sigma}_{\Omega\Omega}$, one calculates the lattice Green
function $\hat{G}(i\omega_n)$ in $\Omega$ by using Eq.\ (\ref{eq12}). 
To avoid double counting of local Coulomb interactions, it is necessary 
to remove at each atomic site in $\Omega$, ${\bf r}_a$, the onsite Coulomb
self-energy term from the lattice Green function. This yields the bath Green
function,
\begin{equation}
\label{eq14}
\hat{G}_{0,aa}(i\omega_n)=\left[\hat{G}_{aa}(i\omega_n)^{-1}
+\hat{\Sigma}_{aa}(i\omega_n)\right]^{-1},
\end{equation}
where $\hat{\Sigma}_{aa}$ is the projection of
$\hat{\Sigma}_{\Omega\Omega}$ on atomic site ${\bf r}_a$ defined by
\begin{equation}
\label{eq15}
\hat{\Sigma}_{aa}(i\omega_n)= \sum_{\tilde{w}, \tilde{w}'}
\langle\tilde{w}\mid\hat{\Sigma}_{\Omega\Omega}(i\omega_n)
\mid\tilde{w}'\rangle c^\dagger_{\tilde{w}}c_{\tilde{w}'},
\end{equation}
with both $\tilde{w}$ and $\tilde{w}'$ located on ${\bf r}_a$.
$\hat{G}_{aa}$, the projection of the lattice Green function $\hat{G}$ on
atomic site ${\bf r}_a$, is defined in the same way.
Both $\hat{\Sigma}_{aa}$ and $\hat{G}_{aa}$ are $N_a\times N_a$ matrices,
where $N_a$ is the number of basis functions centered at ${\bf r}_a$.
Within the single-site approximation, $\hat{\Sigma}_{\Omega\Omega}$ is
diagonal with respect to atomic sites, so that
\begin{equation}
\label{eq16}
\hat{\Sigma}_{\Omega\Omega}(i\omega)=\sum_{a\in\Omega}
\hat{\Sigma}_{aa}(i\omega_n).
\end{equation}

The bath Green function $\hat{G}_{0,aa}$ determines the Weiss mean-field
Hamiltonian at site ${\bf r}_a$. One then adds the local Coulomb interactions
of the form Eq.\ (\ref{eq2}) and solves the single-site many-body impurity 
problem at site ${\bf r}_a$ by numerical methods, such as the quantum Monte
Carlo approach,\cite{Hirsch:86,Georges:92} 
exact-diagonalization\cite{Caffarel:94} (ED), 
or the numerical renormalization group method.\cite{Bulla:08}
The resultant impurity Green function, $\hat{G}^{\rm imp}_{aa}$, is used to derive
the output impurity self-energy via
\begin{equation}
\label{eq17}
\hat{\Sigma}^{\rm imp}_{aa}(i\omega_n)=
\hat{G}_{0,aa}(i\omega_n)^{-1}-\hat{G}^{\rm imp}_{aa}(i\omega_n)^{-1}.
\end{equation}
The key assumption in DMFT is now that this impurity self-energy is a physically
reasonable representation of the lattice self-energy. Thus, 
$\hat{\Sigma}_{aa}(i\omega_n)\approx \hat{\Sigma}^{\rm imp}_{aa}(i\omega_n)$.
This self-energy is therefore used as input in Eqs.\ (\ref{eq13}), (\ref{eq14}) 
in the next iteration. This procedure is repeated until the difference 
between the input and output self-energies becomes sufficiently small for 
all atomic sites in the embedded region $\Omega$.

\section{Results and discussion}
\label{sec_3}

\subsection{Hubbard model}

To demonstrate the DMFT embedding approach we present results for 
the single-band Hubbard model,
\begin{eqnarray}
\hat{H}&=& \sum_{a \sigma} \epsilon_a \hat{n}_{a \sigma}
-\sum_{\langle a b \rangle \sigma}
(t_{ab}\ c^\dagger_{a\sigma} c_{b\sigma}+H.c.)\nonumber \\
&+& \sum_{a} U_a \hat{n}_{a\uparrow}\hat{n}_{a\downarrow},
\label{eq18}
\end{eqnarray}
where $\hat{n}_{a \sigma}=c^\dagger_{a\sigma} c_{a\sigma}$ and the
summation in the second term is taken over nearest neighbor sites.
We consider a simple cubic lattice with its three principal axes
oriented along the $x$, $y$, and $z$ directions. The interface points
in the (001) direction. In each layer, all sites are
assumed to be equivalent (1$\times$1 structure).
We label the site energy of layer $j$ as $\epsilon_j$, the Coulomb energy
of layer $j$ as $U_j$, the $x$ and $y$ components of the inplane transfer
integrals in layer $j$ as $t^x_j$ and $t^y_j$, and the transfer integral
between two nearest-neighbor layers, $j$ and $k$, as $t^z_{jk}$.
The Hamiltonian parameters in $L$ ($R$) represents a particular bulk
crystal with a single atom in the unit cell. The Hamiltonian parameters in
$\Omega$ approach those of the left-hand side (right-hand side) crystal near
the boundary to $L$ ($R$), while they are allowed to deviate from these bulk
parameters in the interior of region $\Omega$. In the present work, we consider
only paramagnetic solutions and omit the spin index $\sigma$ in the discussion
below.

As an input, one needs the embedding potentials of both substrates.
Let us consider the left substrate $L$, whose
site energy, Coulomb energy, and transfer integrals are given by
$\epsilon_L$, $U_L$, $t^x_L$, $t^y_L$, and $t^z_L$.
First, we ignore the Coulomb interactions and derive the embedding potential
for non-interacting electrons as defined by Eq.\ (\ref{eq6a}).
Because of translational symmetry in the plane, the embedding potential is
diagonal with respect to the two-dimensional wave vector ${\bf k}=(k_x, k_y)$ and
can be expressed as $s^L_{ij}({\bf k},\epsilon)$ by introducing a
mixed representation with ${\bf k}$ and layer indices in $\Omega$, $i$ and
$j$. Here, the wave vector is measured in units of the inverse of lattice
constant, i.e., $-\pi \leq k_x, k_y \leq \pi$.
For the present nearest-neighbor transfer model, the only non-vanishing
element is $s^L_{11}({\bf k},\epsilon)$, which is given as
\begin{equation}
\label{eq19}
s^L_{11}({\bf k},\epsilon)=(t^z_{01})^2\
g^L_{00}\left(\epsilon-E_L({\bf k})\right),
\end{equation}
where
\begin{equation}
\label{eq19b}
E_L({\bf k}) = \epsilon_L-2t^x_L \cos(k_x) - 2t^y_L \cos(k_y),
\end{equation}
and $g^L_{jj'}$ with $j,j' <1$ denotes the Green function of a semi-infinite
tight-binding chain with nearest-neighbor transfer integral, $-t^z_L$.
According to Kalkstein and Soven,\cite{Kalkstein:71}
\begin{equation}
\label{eq20}
g^L_{00}(w)=\frac{w-\sqrt{w^2-4 (t^z_L)^2}}{2 (t^z_L)^2}.
\end{equation}

The embedding potential of region $L$ in the presence of Coulomb
interactions is obtained by incorporating the effects of
electron correlations in the bulk crystal into Eq.\ (\ref{eq19}) as
\begin{equation}
\label{eq21}
S^L_{11}({\bf k},i\omega_n)=(t^z_{01})^2\ g^L_{00}(w),
\end{equation}
with
\begin{equation}
\label{eq21b}
w=i\omega_n+\mu-E_L({\bf k})
-\langle a |\hat{\Sigma}_{LL}(i\omega_n)| a \rangle ,
\end{equation}
where $a$ is any site in $L$. On the right-hand side of Eq.\ (\ref{eq21b}),
the Coulomb self-energy in $L$ is determined by a bulk DMFT
calculation. The embedding potential of the right substrate $R$ can be
constructed in the same way.

Using Eq.\ (\ref{eq12}), the lattice Green function in
region $\Omega$ is now  calculated as
\begin{eqnarray}
G_{ab}(i\omega_n)&=&\int \frac{d{\bf k}}{(2\pi)^2}
e^{i{\bf k}\cdot({\bf x}_{mj}-{\bf x}_{m'j'})} \nonumber\\
&\times& \langle j |[i\omega_n+\mu-\hat{H}_{\rm em}({\bf k},i\omega_n)]
^{-1}| j' \rangle,\label{eq22}
\end{eqnarray}
where $a=(m,j)$, $b=(m',j')$, $1\leq j, j'\leq N$. 
In the mixed representation the embedded Hamiltonian,
$\hat{H}_{\rm em}({\bf k},i\omega_n)$, is an $N\times N$ matrix,
\begin{eqnarray}
\langle j|\hat{H}_{\rm em}| j'\rangle&=&E_j({\bf k})
\delta_{jj'}-t^z_{jj'}+\Sigma_j(i\omega_n)\delta_{jj'}\nonumber \\
&+& S^L_{jj'}({\bf k},i\omega_n)+ S^R_{jj'}({\bf k},i\omega_n).
\label{eq23}
\end{eqnarray}
Here, 
\begin{equation}
\label{eq24}
E_j({\bf k}) = \epsilon_j-2t^x_j \cos(k_x) - 2t^y_j \cos(k_y).
\end{equation}
The Coulomb self-energy of layer $j$, $\Sigma_j(i\omega_n)=
\langle a| \hat{\Sigma}_{\Omega\Omega}(i\omega_n)|a\rangle$
with $a=(m,j)$, is diagonal with respect to the layer index
and has no dependency on ${\bf k}$ within the single-site approximation.
As argued above, in a nearest-neighbor tight-binding system, only
the embedding potentials $S^L_{11}({\bf k},i\omega_n)$ and 
$S^R_{NN}({\bf k},i\omega_n)$ are finite. The layer-dependent onsite 
Green function for $a=b=(m,j)$ will be denoted as $G_j(i\omega_n)$
and the corresponding bath Green function as $G_{0,j}(i\omega_n)$.

\subsection{Numerical results}

We consider first the surface of a semi-infinite Hubbard model having
uniform Hamiltonian parameters, i.e., $U_j=U$, $\epsilon_j=-U/2$, and
$t^x_j=t^y_j=t^z_{jk}=t$ for all layers including the surface plane.
By choosing the chemical potential $\mu$ as zero, all layers become
half-filled due to electron-hole symmetry.
For zero temperature, the same system was studied by Potthoff and
Nolting,\cite{Potthoff:99} who showed that there is a uniform critical
Coulomb energy $U_c$, at which both bulk and surface simultaneously undergo
a metal-insulator transition. 
For a complete numerical solution of the DMFT equation,
they adopted a slab geometry consisting of 10 to 20 atomic layers
rather than treating semi-infinite surfaces.

As impurity solver we employ the finite-temperature ED method. Thus, 
for each layer $j$, the bath Green function Eq.\ (\ref{eq14}) is 
projected on to a small cluster consisting of a single impurity 
surrounded by several bath levels. Eq.\ (\ref{eq14}) is therefore  
approximated as
\begin{equation}
\label{eq25}
{G}_{0,j}(i\omega_n)\approx \left(i\omega_n+\mu-E_j
-\sum_{k=1}^{n_b} \frac{|V_{jk}|^2}{i\omega_n-E_{jk}}\right)^{-1},
\end{equation}
where $E_j$ represents an impurity level for layer $j$, $E_{jk}$ the 
corresponding bath levels, and $V_{jk}$ specifies the hybridization matrix. 
We use $n_b=7$ bath orbitals in the numerical results presented below. 
The inclusion of the ficticious impurity level $E_j$ provides a more 
accurate projection of ${G}_{0,j}(i\omega_n)$ than for a cluster
consisting only of bath orbitals. The interacting Green function
of the cluster with onsite Coulomb energy $U$ at finite temperature $T$ is
derived by calculating the low eigenvalues of the cluster via the Arnoldi
algorithm and applying the Lanczos procedure for computing the
excited state Green function. More details of the ED method can be found in
Ref.\ \onlinecite{Perroni:07}.

\begin{figure}[htbp]
\begin{center}
\includegraphics[width=0.4\textwidth]{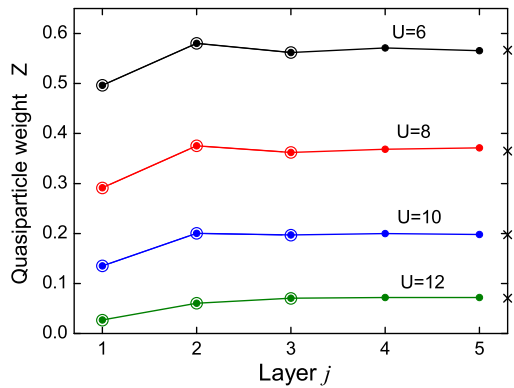}
\end{center}
\caption{\label{Fig1}  (color online)
Quasiparticle weight $Z$ of semi-infinite Hubbard model for simple cubic
lattice in the (001) orientation as a function of layer index $j$.
Temperature $T=0.02$. Solid dots and open circles are results with $N=5$ and $N=3$
embedded layers, respectively.
Crosses on the vertical axis indicate the bulk $Z$ corresponding to four values of
$U$. Lines are drawn as a guide to the eye.}
\end{figure}

In Fig.\ \ref{Fig1} we show the calculated quasiparticle weight $Z$ of
the semi-infinite Hubbard model at $T=0.02$ in the metallic range $U<U_c$
as a function of layer index $j$, where the outermost layer corresponds to
$j=1$. The hopping integral is taken to be $t=1$ and defines the energy scale. 
The crosses on the vertical axis indicate the $Z$ values of the bulk metal
determined by a separate bulk DMFT calculation.  The embedding potential acts on
layer $N$ on the right edge of the surface region. 
Solid dots and open circles provide the results obtained for
$N=5$ and $N=3$ embedded layers, respectively.
The excellent agreement between the two sets of calculations demonstrates
that the embedding potential represents correctly the one-electron
as well as many-body properties of the semi-infinite substrate. Moreover,
it is evident
that one needs only a few embedded layers to simulate the semi-infinite system.
Although the latter point is not crucial for single-band model systems, 
for the calculation of realistic 
multi-orbital materials the embedding treatment yields a substantial
reduction of computer time compared with slab calculations
in which at least $\sim$10 layers must be explicitly taken into consideration.

As can be seen in Fig.~1, the calculated $Z$ exhibits an oscillatory behavior 
near the surface which follows from the Friedel oscillations of the 
layer-dependent density of states. In the first layer, $Z$ is smaller than 
the bulk value, implying that electrons at the surface are more strongly
correlated than in the bulk. As discussed by Potthoff and
Nolting\cite{Potthoff:99} and Liebsch,\cite{Liebsch:03} this is essentially
a one-electron effect arising from the layer dependence of the one-electron
DOS of the cubic tight-binding Hamiltonian.\cite{Kalkstein:71} Because of the
loss of nearest-neighbor sites, the effective band width in the first layer is
reduced, so that Coulomb correlations at the surface are enhanced.

Next, we study the interface between two semi-infinite Hubbard models.
We consider a uniform system with regard to the transfer integrals, i.e.,
$t^x_j=t^y_j=t^z_{jk}=t=1$.
In the left half-space, we choose the Coulomb energy as $U_L=6 $ to
represent a good metal with a relatively large quasiparticle weight, while we
take a variable, larger Coulomb energy $U_R$ in the half-space on the right.
Furthermore, by choosing $\epsilon_j$ as $-U_L/2$ and $-U_R/2$ in the left and
right half-spaces, respectively, and by setting the chemical potential as
$\mu=0$, all layers are half-filled. The same model was recently
investigated by Helmes {\it et al.}\cite{Helmes:08} who used the NRG method 
as impurity solver. 
These authors focused on the critical range for $U_R\sim U_c$
and discussed the scaling behavior of the metallic penetration depth into the Mott
insulator. To reduce finite-size effects, a relatively thick slab
consisting of $\sim$60 layers was used to simulate the interface. 
Also, to avoid numerical difficulties stemming from the energy discretization 
the van Hove singularity of the two-dimensional layer DOS was cut off at a finite
value.  

\begin{figure}[htbp]
\begin{center}
\includegraphics[width=0.4\textwidth]{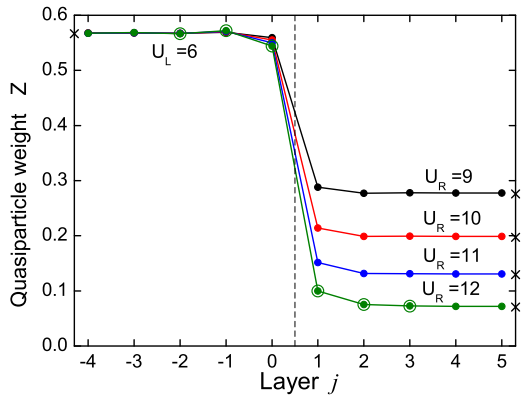}
\end{center}
\caption{\label{Fig2}  (color online)
Quasiparticle weight $Z$ at the interface between two semi-infinite Hubbard
models for simple cubic lattice in (001) orientation as a function of layer
index $j$. The boundary between the two metals is marked by a
dashed line. Temperature $T=0.02 $. Solid dots and open circles show results with
$N=10$ and $N=6$ embedded layers, respectively. Crosses on the vertical axis indicate
bulk $Z$ values. Lines are drawn as a guide to the eye.}
\end{figure}

Figure\ \ref{Fig2} shows the calculated quasiparticle weight
$Z$ in the metallic range $U_R<U_c$ at $T=0.02$ as a function of layer index
$j$, which is measured here relative to the boundary
layer of the left-hand side metal.
To describe the deviation of the electronic structure from that in bulk
metal, we incorporate in the embedded region $N=10$ layers (solid dots), 
of which the left (right) five layers possess Coulomb energy $U_L$ ($U_R$).
For comparison, we also show for $U_R=12$ (open circles) the result obtained 
with only $N=6$ embedded layers. The excellent agreement between the two sets of
calculations corroborates again the efficiency of the embedding method to
treat semi-infinite substrates.

The quasiparticle weight $Z$ of the surface layer of the good metal on the 
left-hand side is seen to be reduced whereas at the surface of the poor metal on 
the right-hand side of the boundary plane it is enhanced. Evidently, the good 
or bad metallic character of one metal spills over into the neighboring metal. 
In contrast to the case of the semi-infinite surface discussed above, 
this is a genuine many-body effect, since the one-electron
DOS is layer-independent if the Coulomb interaction is switched off.
The deviation of $Z$ from the bulk value in poor metal on the right
decreases with the distance from the boundary plane, which is in accord with
the work of Helmes {\it et al}.\cite{Helmes:08}
On the left of the boundary plane, $Z$ is seen to be weakly modified with 
respect to the bulk value only in the first two layers. Thus, as a result of 
better electronic screening in the good metal, $Z$ approaches the bulk 
value more rapidly than in the poor metal. Essentially, one needs to
incorporate only one or two layers in the embedded region to describe the 
interface properties of the good metal on the left-hand side.

\begin{figure}[htbp]
\begin{center}
\includegraphics[width=0.4\textwidth]{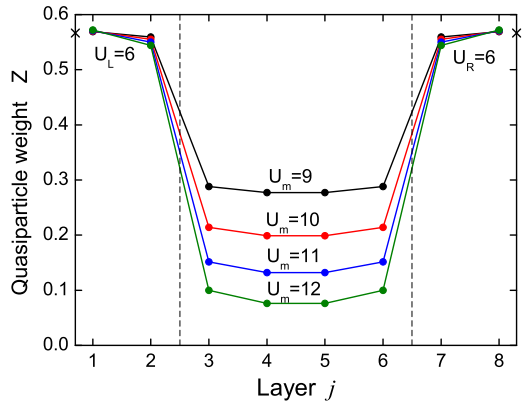}
\end{center}
\caption{\label{Fig3}  (color online)
Quasiparticle weight $Z$ of the heterostructure consisting of a 4-layer
film ($j=3\ldots6$) sandwiched between two semi-infinite Hubbard models for
simple cubic lattice in the (001) orientation.
Crosses on the vertical axes indicate bulk $Z$ values for the left and
right substrates. Temperature $T=0.02$. Lines are drawn as a guide to the
eye.}
\end{figure}

As the third model system, we study a junction in which a finite number of
strongly correlated-electron layers are sandwiched between two weakly correlated 
metals. We adopt again a uniform model with respect to transfer integrals, i.e.,
$t^x_j=t^y_j=t^z_{jk}=t=1$ for all layers. We assign a non-zero but moderate
Coulomb energy $U_{L,R}= 6 $ to both metal substrates,
whereas in the central film we choose a larger Coulomb energy $U_m$.
In Fig.\ \ref{Fig3} we show the calculated quasiparticle weight $Z$ for a
4-layer film as a function of layer index $j$ in the metallic range $U_m<U_c$.
The calculation was carried out using $N=8$ embedded layers, which comprise the
central 4-layer film and the two outermost layers of metal substrates on both
sides. Interestingly, in this thin film $Z$ is very close to that of the
boundary layers of the semi-infinite metal with $U=U_R$ shown in
Fig.\ \ref{Fig2}.
This rapid convergence of $Z$ with increasing film thickness may arise
partly from the peculiarity of the present model in which the one-electron
DOS is layer-independent. Thus, there appear no finite-size effects such as
energy-level discretization in the one-electron spectrum at the junction.

\begin{figure}[htbp]
\begin{center}
\includegraphics[width=0.4\textwidth]{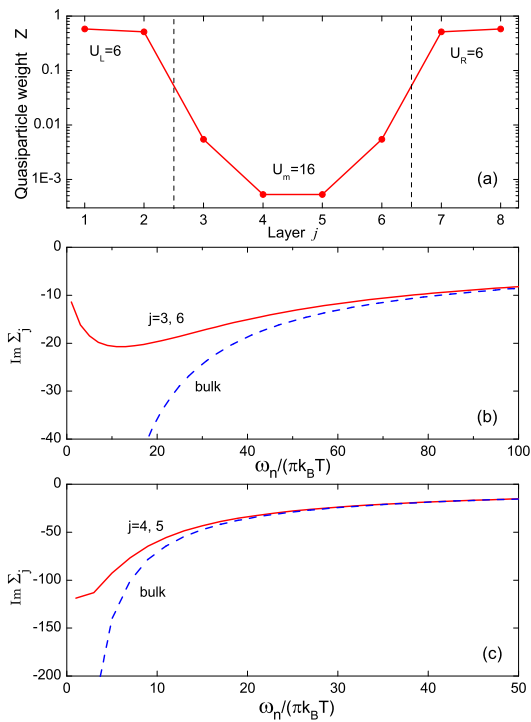}
\end{center}
\caption{\label{Fig4}  (color online)
(a) Quasiparticle weight $Z$ of a 4-layer film with $U_m=16$  sandwiched
between two metals with $U_{L,R}=6$ as a function of layer index $j$.
Lines are drawn as a guide to the eye.
(b) and (c) Imaginary part of the impurity self-energy of boundary layers
($j= 3,\ 6$) and inner layers ($j= 4,\ 5$) of the thin film as a function 
of Matsubara frequency for $U_m=16$ (red solid line). The corresponding bulk 
self-energy at the same Coulomb energy is shown by the blue dashed line.}
\end{figure}

We finally consider a metal/insulator/metal junction.
Figure \ref{Fig4}(a) shows the quasiparticle weight $Z$ of a 4-layer film
with $U_m =16 > U_c$ sandwiched between two metals with $U_{L,R}=6 $
at temperature $T=0.02$.
As discussed by Helmes {\it et al.},\cite{Helmes:08} the metallic states 
decay within the insulating layers, so that $Z$ becomes finite in the film.
In agreement with these NRG results we find that this penetration depth 
within the Mott gap is extremely short. To demonstrate this more clearly, 
we plot in Figs.~4(b) and (c) the imaginary part of the layer-dependent 
Coulomb self-energy $\Sigma_j(i\omega_n)$ for the outer and inner layers of 
the film, respectively, as a function of Matsubara frequency. For comparison, 
we also show the imaginary part of the impurity self-energy in the bulk 
simple-cubic crystal with $U=16$. Whereas the bulk self-energy diverges as 
$\omega_n\rightarrow0$ at this Coulomb energy, the film self-energy tends 
to a finite value because of its contact to the neighboring metal layers.
At the film surface, the finite value is about $-10$, indicating bad metallic
behavior with a rather short electron lifetime and a very small quasi-particle 
weight $Z\approx0.005$. In the second layer, the limiting value of the 
self-energy at low $\omega_n$ is more than one order of magnitude larger
than at the surface, implying correspondingly shorter electronic
lifetime and lower quasi-particle weight $Z$. Thus, apart from a weak 
exponential bad-metallic tail, the Mott gap of the insulating film is
virtually impenetrable.

Electron transport through correlated-electron systems such as oxide
heterostructures and molecules is emerging as an active field of theoretical
studies.\cite{Ferretti:05,Okamoto:08,Thygesen:08}
To our knowledge, previous studies considered only non-interacting metal
leads connected to a central region with Coulomb interactions. It would be
interesting to extend the transport theory to the case of interacting metal
leads as those shown in Figs.\ \ref{Fig3} and \ref{Fig4}.
\bigskip

\section{Summary}
\label{sec_4}
We have presented an efficient embedding scheme for performing DMFT calculations 
for inhomogeneous layered systems such as semi-infinite surfaces and 
heterostructures. In contrast to previous embedding theories based on
tight-binding basis functions, the embedding potential introduced here is 
determined from a separate DMFT calculation for the bulk substrate materials. 
It therefore incorporates not only the one-electron properties but also the 
many-body effects of the semi-infinite systems. The interface region in which
local Coulomb interactions are treated self-consistently via the layer-coupled
DMFT also includes the first few layers of the actual substrates. 
As examples, we have presented 
numerical results for several surfaces and interfaces of the single-band Hubbard 
model. These results demonstrate that the represention of the semi-infinite 
correlated substrates in terms of complex energy-dependent embedding potentials
greatly reduces the numerical effort since only a small number of layers needs
to be explicitly included in the layer-coupled DMFT equation. Thus, the study 
of neutral as well as charged heterostructures involving realistic strongly 
correlated multi-band materials becomes feasible.

\bigskip

\begin{acknowledgments}
One of us (A. L.) would like to thank Theo Costi for comments.
The work of H. I. was supported by the Grand-in-Aid for Scientific Research
(20540191).
\end{acknowledgments}

\end{document}